\documentclass[twocolumn,rotate,aps,prl,amsmath,nobalancelastpage]{revtex4}
\usepackage{amssymb}
\usepackage{graphicx}

\begin{document}

\title{Kolmogorov and Bolgiano scaling in thermal convection: \\
the case of Rayleigh-Taylor turbulence}

\author{G. Boffetta$^{1}$, F. De Lillo$^{1}$, A. Mazzino$^{2}$,
  S. Musacchio$^{3}$}

\affiliation{$^1$Dipartimento di Fisica Generale and INFN, 
Universit\`a di Torino, via P.Giuria 1, 10125 Torino (Italy) \\
$^2$Dipartimento di Fisica, Universit\`a di Genova, INFN and CNISM,
via Dodecaneso 33, 16146 Genova (Italy) \\
$^{(3)}$ CNRS, Lab. J.A. Dieudonn\'e UMR 6621,
Parc Valrose, 06108 Nice (France)}

%\date{\today}
\begin{abstract}
We investigate the statistical properties of Rayleigh-Taylor turbulence 
in a convective cell of high aspect ratio, 
in which one transverse side is much smaller that the others.
We show that the scale of confinement determines the Bolgiano scale 
of the system, which in the late stage of the evolution is characterized 
by the Kolmogorov-Obukhov and the Bolgiano-Obukhov phenomenology 
at small and large scales, respectively.
The coexistence of these regimes is associated to a three to two-dimensional
transition of the system which occurs when the width of the turbulent 
mixing layer becomes larger that the scale of confinement. 
\end{abstract}
%\pacs{PACS?}

\maketitle

%%%%%%%%%%%%%%%%%%%%%%%%%%%%%%%%%%%%%%%%%%%%%%%%%%%%%%%%%%%%%%%%%%%%%%
Turbulent thermal convection appears in many natural phenomena,
from heat transport in stars to turbulent mixing in the atmosphere
and the oceans, and in technological applications \cite{siggia_arfm94,
agl_rmp09,nssd_nat00}.
Turbulent convection is driven by buoyancy forces
generated by temperature fluctuations. These are then mixed
by the turbulent flow itself up to small scales at which 
molecular diffusivity becomes important.
A fundamental problem in thermal convection is the determination
of the statistical properties of velocity and temperature 
fluctuations in the {\it inertial} range of scales in which 
turbulent mixing is at work.

A first step in this direction was done by Obukhov \cite{obukhov_49}
and Corrsin \cite{corrsin_51} who generalized the Kolmogorov 
argument for the statistics of a temperature
field in the so-called passive limit, in which the effects of the 
buoyancy forces on the velocity field are neglected \cite{ss_pra90}.
%In this limit the temperature and velocity fluctuations 
%at scales within the inertial range are both characterized 
%by the same scaling exponent $1/3$.  
%This picture is, {\it a posteriori}, consistent with the 
%assumption that the buoyancy force becomes less and less important at 
%small scales where therefore the passive limit is recovered \cite{ss_pra90}. 
An alternative prediction was proposed by Bolgiano \cite{bolgiano_59}
and Obukhov \cite{obukhov_59}, in discussing the statistics of velocity 
and temperature fluctuations in a stably stratified atmosphere.
The buoyancy forces allow to introduce in the inertial range 
a characteristic scale, the Bolgiano scale $L_B$, 
above which the statistics of the velocity and temperature
is determined by the balance between the buoyancy and inertia forces. 
%This leads to a different prediction for the scaling exponents, 
%namely $3/5$ for the velocity and $1/5$ for the temperature field. 
In spite of many years of experimental and numerical investigations,
no clear evidence of this scenario has been given \cite{lx_arfm10}. 

In this Letter we show that in three-dimensional 
Rayleigh-Taylor turbulent convection
the Bolgiano scale is determined by the geometrical scales of the
setup. By confining the flow in a box with one dimension (e.g. $y$) 
much smaller than the other two, the scale $L_y$ becomes the
Bolgiano scale of the system. By means of state of the art numerical 
simulations
of primitive equations we find coexistent Kolmogorov-Obukhov scaling
at scales smaller than $L_y$ and Bolgiano scaling at scales 
larger than $L_y$. Our geometrical interpretation of the Bolgiano 
scale suggests a new direction for numerical and experimental 
investigations of scaling properties in thermal convection.

%------------------------------------------------------------------------
\begin{figure}[htb!]
\includegraphics[clip=true,keepaspectratio,width=7.5cm]{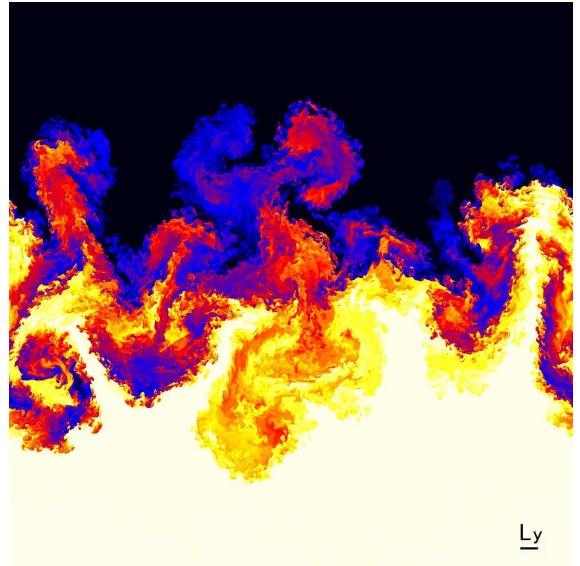}
\caption{Central part of a vertical $(x,z)$ section of the temperature field 
(white=hot, black=cold)
from a direct numerical simulations of Boussinesq equations (\ref{eq:1})
at resolution $4096 \times 128 \times 8192$ taken at time
$t=3.0 \tau$ (see text for the definition of $\tau$).
Parameters are $\beta g=0.5$,
$\nu=\kappa=2 \times 10^{-5}$ and the initial temperature jump is $\theta_0=1$.
The picture clearly shows that large-scale 2D structures coexist 
with small-scale 3D turbulence. The segment represents the  
transverse size $L_y$.}
\label{fig1}
\end{figure}
%------------------------------------------------------------------------

%%%%%%%%%%%%%%%%%%%%%%%%%%%%%%%%%%%%%%%%%

Rayleigh-Taylor (RT) turbulence is one of the simplest configurations
of thermal convection in which a cold, heavier layer of fluid is placed 
on the top of an hot, lighter layer in a gravitational field.
Rayleigh-Taylor instability occurs in several phenomena ranging
from geophysics, to astrophysics to technological applications
\cite{mammatus06,cc_natphys06,imsy_nat05}.
The gravitational instability develops in an intermediate layer of 
turbulent fluid (the mixing layer) the width of which grows in time.  

%------------------------------------------------------------------------
\begin{figure}[htb!]
  \includegraphics[clip=true,keepaspectratio,width=8.5cm]{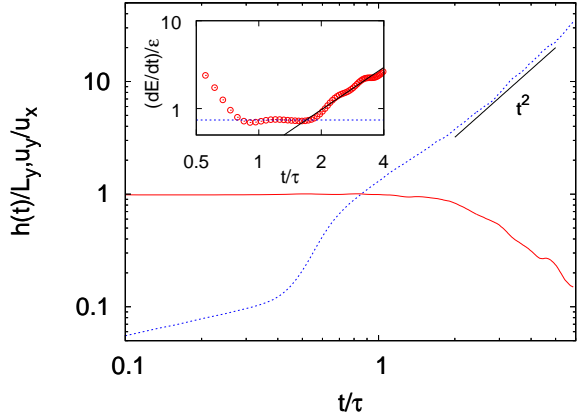}
\caption{Temporal evolution of the mixing layer width $h(t)$ (blue, dashed line)
and the ratio of $y$ component to $x$ component of rms velocity fluctuations
(red, solid line). 
Inset: 
Ratio between kinetic energy growth rate $dE/dt$ and energy
dissipation rate $\varepsilon$ as a function of time. The horizontal
line indicates a constant ratio $\sim 0.75$ and the continuous line
represents the scaling $t^{8/5}$.}
\label{fig2}
\end{figure}
%------------------------------------------------------------------------

We consider miscible RT turbulence at low Atwood numbers.
Within the Boussinesq approximation, 
the equations for the dynamics of the velocity field ${\bf u}$ 
coupled to the temperature field $T({\bf x},t)$ 
(which is proportional to the density $\rho$ via the thermal expansion 
coefficient $\beta$ as $\rho=\rho_0 [1-\beta (T-T_0)]$ where 
$\rho_0$ and $T_0$ are reference values) read: 
\begin{eqnarray}
\partial_t {\bf u} + {\bf u} \cdot {\bf \nabla} {\bf u} & = & 
- (1/\rho_0) {\bf \nabla} p + \nu \nabla^2 {\bf u} - 
\beta {\bf g} T \nonumber \\
\partial_t T + {\bf u} \cdot {\bf \nabla} T & = &  \kappa \nabla^2 T 
\label{eq:1} 
\end{eqnarray}
together with the incompressibility condition 
${\bf \nabla} \cdot {\bf u} = 0$.
In (\ref{eq:1}) ${\bf g}=(0,0,-g)$ is gravity acceleration, 
$\nu$ is the kinematic viscosity and $\kappa$ is the thermal diffusivity. 
The initial condition for the RT problem is an unstable temperature
jump $T({\bf x},0)=-(\theta_0/2)sgn(z)$ in a fluid at rest
${\bf u}({\bf x},0)=0$.

As the system evolves, the available potential energy 
$P=-\beta g \langle z T \rangle $ is converted into 
kinetic energy $E=(1/2) \langle |{\bf u}|^2 \rangle$ 
at a rate that can be estimated from the energy balance: 
\begin{equation}
-{dP \over dt} =\beta g \langle w T \rangle = {dE \over dt} + \varepsilon
\label{eq:2}
\end{equation}
where $w$ is the vertical velocity and 
$\varepsilon = \nu \langle (\partial_{\alpha} u_{\beta})^2 \rangle$
is the viscous energy dissipation rate.
From the dimensional balance between the loss of potential energy 
and the increase of kinetic energy one has that 
typical velocity fluctuations grow as $u_{rms} \simeq \beta g \theta_0 t$, 
and therefore the width of the turbulent mixing layer $h(t)$, 
shown in Fig.~\ref{fig2},  
grows following the accelerated law $h(t) \simeq \beta g \theta_0 t^2$.
The integral scale $L(t)$ of the turbulent flow,  
defined as the largest scale on which kinetic energy is injected, 
is expected to grow proportionally to the geometrical scale $h(t)$
\cite{bmmv_pof10}. 

According to the phenomenological theory of RT turbulence, 
developed in \cite{chertkov_prl03}, 
the scaling behavior of the velocity and temperature fluctuations
in the range of scales between the integral scale $L$ 
and the dissipative scale $\eta$ strongly depends
on the dimensionality of the flow. 

For the three-dimensional (3D) case, 
one assumes that the buoyancy force $\beta {\bf g} T$ balances 
the inertia term in (\ref{eq:1}) at the integral scale $L(t)$ 
and becomes negligible as the cascade proceeds towards small scales, 
consistently with the Kolmogorov--Obukhov phenomenology. 
For velocity and temperature fluctuations $\delta u(r)=u(x+r)-u(x)$ 
($u$ denoting one velocity component)
and $\delta T(r)=T(x+r)-T(x)$ one therefore expects \cite{chertkov_prl03}
\begin{equation}
\begin{array}{l}
\delta u(r) \simeq \varepsilon^{1/3} r^{1/3} \\
\delta T(r) \simeq  \varepsilon_T^{1/2} \varepsilon^{-1/6} r^{1/3} 
%\delta u(r) \simeq (\beta g \theta_0)^{2/3} t^{1/3} r^{1/3} \\
%\delta T(r) \simeq \theta_0 (\beta g \theta_0)^{-1/3} t^{-2/3} r^{1/3} 
\end{array}
\label{eq:4}
\end{equation}
where the energy dissipation rates $\varepsilon$ grows in time as 
$\varepsilon(t) \simeq (\beta g \theta_0)^2 t$, 
following adiabatically the dynamics of the large eddies, 
while the temperature dissipation rates decreases as 
$\varepsilon_T(t) \simeq \theta_0^2 t^{-1}$ 
\cite{cc_natphys06,bmmv_pre09}.

This scenario is not consistent in two dimensions (2D) where
kinetic energy is transferred toward large scales developing an 
inverse cascade \cite{km_rpp80}.
In this case the buoyancy term injects energy at all scales generating
a non-constant-in-wavenumber energy flux. 
As a consequence, velocity and temperature fluctuations follow the 
Bolgiano-Obukhov scaling \cite{chertkov_prl03}
\begin{equation}
\begin{array}{l}
\delta u(r) \simeq \varepsilon_T^{1/5} (\beta g)^{2/5} r^{3/5} \\
\delta T(r) \simeq \varepsilon_T^{2/5} (\beta g)^{-1/5} r^{1/5}
%\delta u(r) \simeq (\beta g \theta_0)^{2/5} t^{-1/5} r^{3/5} \\
%\delta T(r) \simeq \theta_0 (\beta g \theta_0)^{-1/5} t^{-2/5} r^{1/5}
\end{array}
\label{eq:5}
\end{equation}
which has been verified in numerical simulations of 2D RT turbulence
\cite{cmv_prl06}.

%The Bolgiano scale $L_B$ is defined as the scale at which inertial 
%and buoyancy terms balance. From the above considerations one has
%therefore that in Rayleigh-Taylor convection $L_B$ coincides with 
%the largest scale of the inertial range in 3D and with the
%smallest scale in 2D.

%%%%%%%%%%%%%%%%%%%%%%%%%%%%%%%%%%%%%%%%%
%------------------------------------------------------------------------
\begin{figure}[t!]
  \includegraphics[clip=true,keepaspectratio,width=8.5cm]{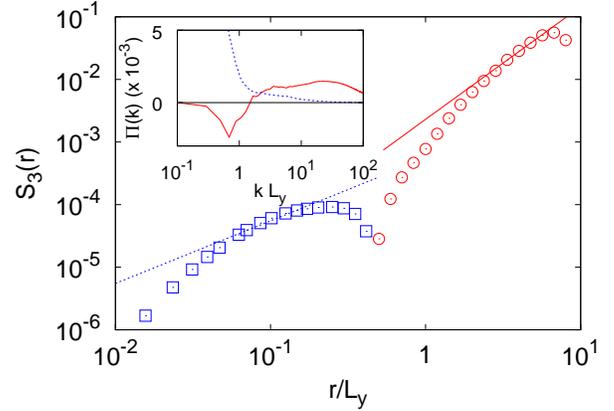}
\caption{Third order longitudinal velocity structure function $S_3(r)$ 
(red circles) and $-S_3(r)$ (blue squares) computed in the mixing 
layer. The two lines represent
Kolmogorov scaling $r$ (blue dotted) and Bolgiano scaling $r^{9/5}$
(red continuous). Inset: contribution to the energy flux in Fourier 
space $\Pi(k)$ from the nonlinear term (red, continuous line) and 
from the buoyancy term (blue, dashed line).  }
\label{fig3}
\end{figure}
%------------------------------------------------------------------------

Let us consider now a convective cell with large aspect
ratio $L_x,L_z \gg L_y$. 
At short times, when $L(t)<L_y$ the turbulent flow in the mixing layer 
can be considered three-dimensional and a direct cascade with 
Kolmogorov-Obukhov scaling (\ref{eq:4})
is expected. At later times, when $L(t)>L_y$ the flow cannot be fully
three dimensional at large scales, as a consequence of the geometrical 
constrain in the $y$ direction. 
The question arises whether the two-dimensional phenomenology 
appears at scales $L_y < r < L(t)$. For this to happen most of the
injected by buoyancy forces should go to large scales producing an
inverse cascade with Bolgiano-Obukhov scaling (\ref{eq:5}).
However, a residual direct cascade to scales $r < L_y$ should 
be present with a flux
$\varepsilon(t)$ given by the matching of the scaling
(\ref{eq:4}) and (\ref{eq:5}) at $r=L_y$:
%We expect that at scales $L_y < r < L(t)$ the statistics of the turbulent flow 
%becomes effectively two-dimensional with inverse energy cascade and 
%Bolgiano-Obukhov scaling exponents (\ref{eq:5}). 
%In this regime most of the energy injected by buoyancy forces 
%goes to large scales but a residual direct cascade
%to scales $r < L_y$ is present with a flux 
%$\varepsilon(t)$ given by the matching of the scaling 
%(\ref{eq:4}) and (\ref{eq:5}) at $r=L_y$:
\begin{equation}
\varepsilon(t) \simeq (\beta g \theta_0)^{6/5} L_y^{4/5} t^{-3/5} 
\label{eq:6}
\end{equation}
The time of the transition from 3D to 2D behavior is given by
continuity requirement in energy dissipation, i.e. equating
(\ref{eq:6}) with 3D dissipation $(\beta g \theta_0)^2 t$ which
gives $\tau=(L_y/(\alpha \beta g \theta_0))^{1/2}$ where 
$\alpha \simeq 0.02$ is a dimensionless number obtained from
numerical simulations \cite{bmmv_pre09}.

Summarizing, the long-time behavior of RT turbulence with large aspect
ratio is the following. 
A small scales $\eta \le r \le L_y$ 
%($\eta(t) \sim t^{3/20}$ is the Kolmogorov scale) 
a three-dimensional direct cascade with Kolmogorov-Obukhov scaling is expected. 
At large scales $L_y \le r \le L(t) \sim t^2$ 
a two-dimensional inverse cascade with Bolgiano-Obukhov scaling 
should be observed. 
%We remark that the presence of simultaneous inverse 
%and direct cascades has been recently detected in standard hydrodynamic 
%turbulence \cite{cmv_prl10}.

The above predictions have been tested against the results 
of state of the art, high resolution numerical simulations 
of equations (\ref{eq:1}), is a large aspect ratio geometry 
with $L_z/L_x = 2$, 
$L_y/L_x =1/32$, in which the flow thus results strongly confined in the $y$
direction. The integration of equations (\ref{eq:1}), 
discretized on a $4096 \times 128 \times 8192$ grid 
with periodic boundary conditions, has been performed with 
a fully parallel pseudospectral code, with $2/3$-dealiasing,
running on a IBM-SP6 supercomputer. 
%The instability is seeded by perturbing the initial interface
%with a superposition of small amplitude waves in a narrow range
%of wavenumbers around the  most unstable linear mode \cite{rda_jfm05}.

Figure~\ref{fig1} shows a vertical section
of the temperature field in the late stage of the 
simulations. Large scales, 2D structures are clearly observed.

As shown in Figure~\ref{fig2}, at $t \simeq \tau$, 
when the mixing layer scale $h(t)$ 
becomes of the order of $L_y$ 
the flow becomes increasingly anisotropic with $u_y \ll u_x, u_z$.
In this conditions, we observe a transition from 3D to 2D turbulent 
behavior, clearly signaled by a change in the ratio 
between the energy growth rate $dE/dt$ and the viscous dissipation 
rate $\varepsilon$ (the energy flux to small scales in the direct 
cascade).
In the 3D regime both these quantities grow linearly in time and
therefore their ratio is constant. 
After the transition the inverse cascade sets in and 
$(dE/dt)/\varepsilon \sim t^{8/5}$,
as follows from (\ref{eq:2}) and (\ref{eq:6}). Both behaviors are 
evident in Fig.~\ref{fig2}.

%------------------------------------------------------------------------
\begin{figure}[t!]
\includegraphics[clip=true,keepaspectratio,width=8.5cm]{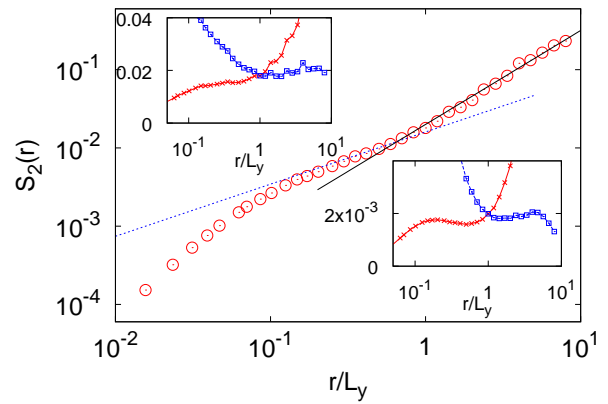}
\caption{Second-order longitudinal velocity structure function $S_2(r)$.
The two lines represent Kolmogorov scaling $r^{2/3}$ (blue dotted) 
and Bolgiano scaling $r^{6/5}$ (red continuous).
Upper inset: Second-order structure function $S_2(r)$ compensated
with Kolmogorov (red crosses) and Bolgiano (blue squares) scaling.
Lower inset: Fourth-order structure function $S_4(r)$ compensated
with Kolmogorov (red crosses) and Bolgiano (blue squares) scaling.}
\label{fig4}
\end{figure}
%------------------------------------------------------------------------

In the late stage of the evolution, when $L(t) > L_y$ we expect the
simultaneous presence of a direct and an inverse cascade in the
two range of scales  $r< L_y$ and $r> L_y$, respectively. 
This can be verified by computing the scale
dependent energy flux, given by the third-order structure function
of longitudinal velocity increments (i.e. taken along the local 
velocity direction)
$S_3(r) \equiv \langle (\delta_{\parallel} u(r))^3 \rangle$.
For isotropic three dimensional turbulence, the classical result 
due to Kolmogorov predicts \cite{frisch_95} $S_3(r)=-(4/5) \varepsilon r$.

As shown in Fig.~\ref{fig3}, at small scales $r<L_y$ $S_3(r)$ is
negative and, in a narrow range of scales, compatible with 
the Kolmogorov prediction $S_3(r) \sim r$.
At scales $r>L_y$, $S_3(r)$ becomes positive, 
signaling the reversal of the energy cascade, 
and displays a scaling behavior $S_3(r) \sim r^{9/5}$
consistent with (\ref{eq:5}).

The inset of Fig.~\ref{fig3} shows the contributions
of the inertia and buoyancy terms to the energy flux in Fourier 
space $\Pi(k) \equiv (d/dt) \int_{k}^{\infty} E(p) dp$
where $E(p)$ is the energy spectrum and time derivative is computed
by taking into account, separately, 
the non-linear and buoyancy terms of (\ref{eq:1}). 
Al low wavenumbers $kL_y < 1$ the buoyancy contribution is dominant, 
and the negative sign of the inertial contribution to the 
energy flux confirms the presence of a 2D inverse cascade. 
At high wavenumbers $kL_y > 1$ the buoyancy contribution
becomes sub-dominant, and one recovers a constant positive flux 
characteristic of the 3D regime \cite{frisch_95}. 

%------------------------------------------------------------------------
\begin{figure}[t!]
\includegraphics[clip=true,keepaspectratio,width=8.5cm]{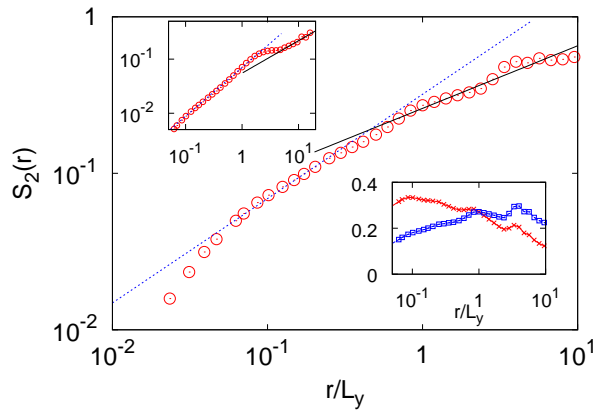}
\caption{Second-order temperature structure function $S_2(r)$.
The two lines represent Kolmogorov scaling $r^{2/3}$ (blue dotted) 
and Bolgiano scaling $r^{2/5}$ (red continuous).
Lower inset: Second-order structure function $S_2(r)$ compensated
by Kolmogorov (red crosses) and Bolgiano (blue squares) scaling.
Upper inset: Fourth-order temperature structure function together
with power laws corresponding to best fit scaling at small scales
$r^{0.9}$ (blue dotted) and at large scales $r^{0.6}$ (red
continuous).}
\label{fig5}
\end{figure}
%------------------------------------------------------------------------

The coexistence of Kolmogorov-Obukhov scaling at small scales 
and Bolgiano-Obukhov scaling at large scales is confirmed by 
behavior of the structure functions of 
longitudinal velocity increments 
$S_p(r) \equiv \langle (\delta_{\parallel} u(r))^p \rangle$
and of temperature increments 
$S^T_p(r) \equiv \langle (\delta T(r))^p \rangle$, 
shown in Figures~\ref{fig4} and~\ref{fig5}.
The transition between the two regimes occurs at the 
Bolgiano scale which is found to be $r \simeq L_y$. 

In three dimensional turbulence, small deviations from the dimensional 
predictions are expected in the scaling of the velocity fluctuations, 
indicating the presence of small scale intermittency \cite{frisch_95}.
Here, we found much stronger corrections in the 
statistics of temperature fluctuations, whose $4$-th order 
structure function strongly 
differs from the dimensional scaling at small scales with a 
best fit exponent $0.9$ close to the corresponding exponent for
passive scalar in three-dimensional turbulence.
At large scale, temperature structure functions show strong fluctuations
due to the presence of regions of unmixed fluid within the mixing layer,
as shown in Fig.~\ref{fig1}. Nonetheless, a very short range of scaling
also for the $4$-th order structure function is observed with a scaling
exponent close to the intermittent value $0.6<4/5$ measured in pure
$2D$ RT simulations \cite{cmv_prl06}.
%These regions correspond to 2D 
%structures and are independent on the transverse $y$ direction
%and therefore contaminate the temperature statistics of a single run
%already at the second order, as shown in Fig.~\ref{fig5}.
%Indeed, ensemble average over many realizations of the flow is
%required to have statistical convergence at large scales
%\cite{cmv_prl06}.

%%%%%%%%%%%%%%%%%%%%%%%%%%%%%%%%%%%%%%%%%%%%%%%%%%%%%
% conclusioni 

Our numerical findings supports the phenomenological prediction that in RT
convection the Bolgiano scale is determined by the aspect ratio of the
convective cell. The presence of Bolgiano-Obukhov scaling is associated to a
dimensional transition of the flow which occurs when the width of the mixing
layer becomes larger than the confining scale $L_y$ 
\cite{cmv_prl10,sbx_prl10}.
This poses the intriguing
question on whether in generic convective systems the Bolgiano-Obukhov
phenomenology could be observed whenever the turbulent flow is confined by
geometrical constraints and/or physical mechanisms (such as rotation or
stratification) in convective cells with small aspect ratio.
Indeed, recent experiments in soap film convection observe Bolgiano
scaling for large values of the Rayleigh number \cite{zw_prl05,sipk_prl10}. 
Despite the similarities with our results, we remark the presence of
important differences, as the experimental data show both the presence
of intermittency in the velocity field \cite{zw_prl05} and the absence
of intermittency in the temperature field \cite{sipk_prl10}. 
On the contrary, our simulations shows the absence of intermittency
for the velocity field which performs an inverse cascade and are
compatible with some intermittency for the temperature field in
the direct cascade.
It would be therefore extremely interesting to have a direct comparison of
experimental and numerical data on thermal convection in quasi-two-dimensional
flow which would give new insights on the fundamental issue of 
intermittency in turbulent convection.

%Moreover, our results demonstrate the possibility to generate a 
%two-dimensional RT laboratory experiment in a three dimensional fluid without 
%the need of an artificial mechanical forcing, because the system is naturally 
%driven by buoyancy. 

Numerical simulations have been done at the Cineca 
under the HPC-EUROPA2 project (project number: 228398) 
with the support of the European Commission - Capacities Area - 
Research Infrastructures.
We thank G. Erbacci and the staff at Cineca Supercomputing Center 
(Bologna, Italy) for their support. 

%%%%%%%%%%%%%%%%%%%%%%%%%%%%%%%%%%%%%%%%%%%%%%%%%%%%%%%%%%%%%%%%%%%%
\bibliography{biblio}{}

\end{document}